# Sensing Models and Its Impact on Network Coverage in Wireless Sensor Network

Ashraf Hossain, P. K. Biswas
Department of Electronics & Electrical Communication Engineering, Indian Institute of Technology Kharagpur, Kharagpur, West Bengal-721302, India
E-mail: {ashraf, pkb}@ece.iitkgp.ernet.in

S. Chakrabarti
G. S. Sanyal School of Telecommunications, Indian Institute of Technology Kharagpur, Kharagpur, West Bengal-721302, India
E-mail: saswat@ece.iitkgp.ernet.in

*Abstract*—Network coverage of wireless sensor network (WSN) means how well an area of interest is being monitored by the deployed network. It depends mainly on sensing model of nodes. In this paper, we present three types of sensing models viz. Boolean sensing model, shadow-fading sensing model and Elfes sensing model. We investigate the impact of sensing models on network coverage. We also investigate network coverage based on Poisson node distribution. A comparative study between regular and random node placement has also been presented in this paper. This study will be useful for coverage analysis of WSN.

*Keywords- sensing model; deterministic sensing model; probabilistic sensing model; random deployment; regular deployment; network coverage*

## I. INTRODUCTION

Wireless sensor network (WSN) consists of a large number of energy-constrained nodes that are deployed for monitoring multiple phenomena of interest. A sensor node consists of a sensing unit, a processing unit, a radio transceiver and a power management unit [1]. Sensor nodes produce some measurable responses to the changes in physical or chemical conditions and transmit these responses to a common sink over a wireless channel. The nodes in a wireless sensor network are generally energy-constrained, as the battery of a node may not be recharged.

Network coverage is an important issue for WSN. It means how well an area of interest is being monitored by a network. Usually, a node has a limited sensing range. Any event is said to be detectable if at least one node lies within its observable range. A node will cover less area when it is placed near the boundary of the area of interest than when it is placed at the central zone. This is known as boundary or border effect. The reason is that some portion of its sensing area will lie outside the area of interest when it is placed near the border [2].

Coverage has been studied for WSN by several authors [2], [3], [4]. All the reported work considers only the Boolean sensing model. Very recently, Tsai [5] has studied sensing coverage for randomly deployed wireless sensor network in shadow-fading environment. In [5], the Boolean sensing model and shadow-fading sensing model have been considered for analyzing network coverage. Besides these there is another reported sensing model in the literature known as Elfes sensing model [6]. In [7], the authors have studied the network coverage for Elfes sensing model. In this paper, we investigate the impact of sensing model on the network coverage. Both the deterministic and probabilistic sensing models have been investigated for coverage analysis of WSN. We have presented a comparative study of network coverage among the different sensing models to show the impact of sensing models on network coverage.

The rest of the paper is organized as follows. Section II presents the system model and sensing models. Network coverage is presented in section III. Coverage analysis for regular placement has been provided in section IV. Section V presents numerical results. Finally, section VI concludes the paper

## II. SYSTEM MODEL AND SENSING MODELS

We consider an area of interest $A$ where $N$ nodes are randomly distributed. Let us assume that the nodes are uniformly deployed with homogeneous node density $\rho = N/A$.

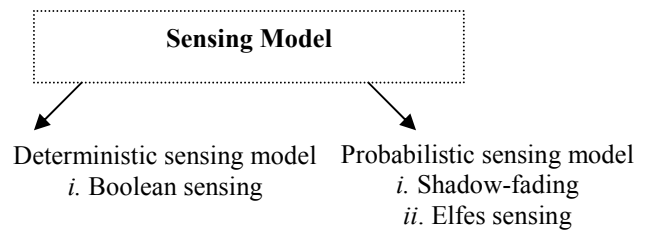

Fig. 1. Sensing models of node: A taxonomy

Two detection models are reported in the literature: individual detection model and cooperative detection model.

For simplicity we assume the first one where each sensor independently detects an event. In the individual detection model, a node detects an event if the received signal strength is greater than the threshold value of detection, known as the sensing sensitivity. The detection process depends on the strength of the emitted signal, behavior of the environment and the hardware of the node. There are two types of sensing models viz. deterministic sensing model and probabilistic sensing model (Fig. 1). Boolean sensing model falls under deterministic category while shadow-fading sensing model and Elfes sensing models fall under probabilistic category. In this section, we present all the three sensing models.

*A. Boolean sensing model*–If the occurrence of the event is within the sensing range of a node then the event will be assumed to be detected, otherwise not. This model ignores the dependency of the condition of the environment (obstacles such as building, foliage) and the strength of the emitted signal on the task of sensing. Usually, the area covered by a sensor node is a circle with radius equals to sensing radius of the node.

*B. Shadow-fading sensing model*–The dependency of all the factors (obstacles such as building, foliage) have been taken into account in this sensing model. Here, the sensing ability of a node is not uniform in all the directions. This is similar to shadowing in radio wave propagation. Assuming log-normal shadowing path loss model, the probability that an event at a distance $x$ from the node will be detected is given by [5]

$$P_{det}(x) = Q\left(\frac{10n \log_{10}(x/r_s)}{\sigma}\right) \quad (1)$$

where, $Q(x) \triangleq \frac{1}{\sqrt{2\pi}} \int_x^\infty e^{-y^2/2} dy$,

$n$ is the path loss exponent ($2 \leq n \leq 4$) [8],
$r_s$ is the sensing radius with out fading,
and $\sigma$ is the fading parameter.

*C. Elfes sensing model*– According to this model [6], the probability that a sensor detects an event to a distance $x$ is

$$p(x) = \begin{cases} 1, & x \leq R_1 \\ e^{-\lambda(x-R_1)^\gamma}, & R_{max} > x > R_1 \\ 0, & x \geq R_{max} \end{cases} \quad (2)$$

where, $R_1$ defines the starting of uncertainty in sensor detection and the parameters $\lambda$ and $\gamma$ are adjusted according to the physical properties of the sensor. $R_{max}$ is the maximum sensing range of the node. This model is more general because it becomes Boolean sensing model when $R_1 = R_{max}$. Equation (2) can be approximated for $R_1 = 0$ and $\gamma = 1$ as

$$p(x) = \begin{cases} e^{-\lambda x}, & R_{max} > x \geq 0 \\ 0, & x \geq R_{max} \end{cases} \quad (3)$$

In the next section, we derive the network coverage for different sensing models.

### III. NETWORK COVERAGE

The network coverage is defined as the ratio of covered area by the network to the area of interest. It depends on the sensing model, number of nodes, node placement strategy. We present how the sensing model affects network coverage in WSN.

*A. Network coverage for Boolean sensing model*

Assume $r_s$ and $A$ are the sensing radius and area of interest respectively. Any event in $A$ will be detected by any arbitrary sensor if it is within $r_s$ distance from the event. The probability that the event will be detected by an arbitrary sensor is $p = \pi r_s^2/A$ (neglecting boundary effect). The event will be undetected by the arbitrary sensor is equal to $(1-p)$. $N$ sensor nodes are deployed randomly. Thus the probability that the event will not be detected by any one of the node is $P_{undet} = (1-p)^N$. The probability that the event will be detected by at least one of the $N$ nodes is equal to the coverage fraction and is

$$f_a = 1 - P_{undet} = 1 - (1-p)^N \quad (4)$$

Equation (4) can also be approximated as $f_a = 1 - exp(-Np)$.
Here, we have neglected the boundary effect. The network coverage for shadow-fading is given in [5]. Now, we present the network coverage in the light of Elfes sensing model.

*B. Network coverage for Elfes sensing model*

We assume that nodes are randomly deployed over an area $A$ and one such sensor is at a distance $x$ from the event (Fig. 2). The probability that a specific node is deployed at a location with a distance $x$ to the event is $2\pi x dx/A$, where $dx$ is a small increment in distance $x$. The probability that the event is sensed by the sensor is $P_{det} = \frac{1}{A} \int_{x=0}^{R_{max}} p(x) 2\pi x dx$.

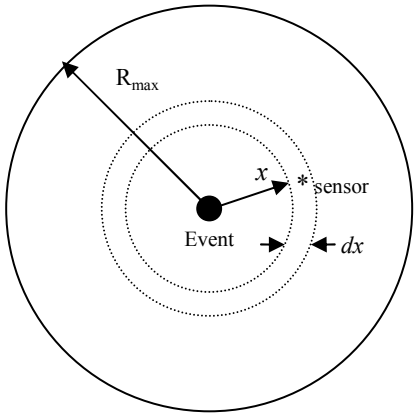

Fig. 2. The sensing for probabilistic model

For $\gamma = 1$ in (2), $P_{det}$ reduces to

$$P_{det} = \frac{\pi R_1^2}{A} + \frac{2\pi}{A\lambda^2}\left[(1+\lambda R_1) - e^{-\lambda(R_{max}-R_1)}(1+\lambda R_{max})\right] \quad (5)$$

The coverage fraction can be found from (4) where $p$ will be replaced by $P_{det}$ and can be simplified for $R_1 = 0$ as

$$f_a = 1 - \exp\left[\frac{2\pi N}{A\lambda^2}\left\{(\lambda R_{max}+1)e^{-\lambda R_{max}} - 1\right\}\right] \quad (6)$$

Equation (6) provides the necessary formula for studying coverage based on probabilistic sensing model. The network coverage is dependent on node sensing model. The expression of network coverage is very simple for Boolean sensing model. It is little bit complex for probabilistic sensing model.

Equation (6) can also be derived using Poisson node distribution. According to Poisson distribution, the probability mass function is

$$\Phi(k) = \frac{e^{-\rho a}(\rho a)^k}{k!}, \quad k = 0, 1, 2, \ldots \quad (7)$$

where, $\rho$ is the node density and $k$ is the number of nodes reside in an area of $a$.

Considering Fig. 2, the probability that the event will not be detected by the nodes over the annular ring of thickness $dx$ is

$$\begin{aligned} P_{un\,det}(x) &= \Phi(0)[1-p(x)]^0 + \Phi(1)[1-p(x)]^1 \\ &\quad + \Phi(2)[1-p(x)]^2 + \ldots + \Phi(N)[1-p(x)]^N \\ &= \sum_{k=0}^{N} \Phi(k)[1-p(x)]^k \quad (8) \\ &= \sum_{k=0}^{N} \frac{e^{-\rho 2\pi x dx}(\rho 2\pi x dx)^k}{k!}[1-p(x)]^k \\ &= e^{-\rho 2\pi x dx} \sum_{k=0}^{N} \frac{(\rho 2\pi x dx)^k}{k!}[1-p(x)]^k \end{aligned}$$

For Poisson node distribution, the assumption $N \to \infty$ gives a closed form expression for (8)

$$P_{un\,det}(x) = e^{-\rho 2\pi x p(x) dx} \quad (9)$$

Probability that the target location is not sensed by the network is

$$\begin{aligned} P_{ns} &= \prod_{x=0}^{R_{max}} P_{un\,det}(x) \\ &= \prod_{x=0}^{R_{max}} e^{-\rho 2\pi x p(x) dx} \quad (10) \\ &= e^{-\sum_{x=0}^{R_{max}} 2\pi \rho x p(x) dx} \end{aligned}$$

Using Riemannian notation we can write the summation by integration

$$P_{ns} = e^{-\int_0^{R_{max}} 2\pi \rho x p(x) dx} \quad (11)$$

Using (3), we get from (11)

$$P_{ns} = \exp\left[\frac{2\pi \rho}{\lambda^2}\left\{(\lambda R_{max}+1)e^{-\lambda R_{max}} - 1\right\}\right] \quad (12)$$

The coverage fraction can be found from $f_a = 1 - P_{ns}$ and it is same as (6).

## IV. NODE PLACEMENT AND COVERAGE ANALYSIS

In a WSN nodes can be deployed randomly or regularly. There is a trade-off between these in terms of number of nodes, deployment cost, deployment time and feasibility of placement scheme. In this section, a comparative study between the regular and random node placement has been presented in connection with network coverage.

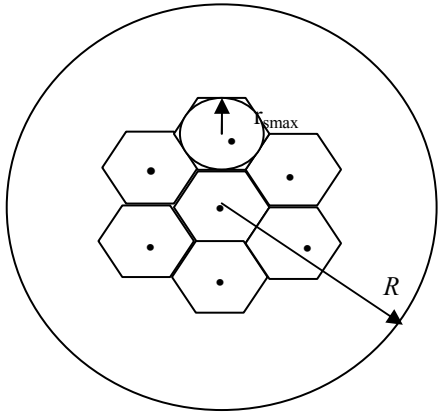

Fig. 3. Scheme of regular node placement

Let us consider regular node placement. We divide the area of interest, $A$ into regular hexagonal cells (Fig. 3). We assume that nodes are placed at the centre of the hexagonal cells and each cell contains only one node. Now, we impose a condition that the node has variable sensing radius $r$ and it varies between 0 and $r_{smax}$, where $r_{smax}$ is the radius of the inscribed circle of the hexagonal cell. This assumption may be considered as the consequence of probabilistic sensing model. Here, we are interested in determining the coverage fraction for regular placement of nodes. For, $R \gg r_{smax}$, the coverage fraction can be approximated as

$$f_a = \frac{\pi}{2\sqrt{3}}\left(\frac{r}{r_{s\max}}\right)^2 \quad (13)$$

The coverage fraction attains the maximum value (90.69%) for $r = r_{smax}$. The number of hexagonal cells, $N_h$ can be expressed as

$$N_h = 0.9069 (R/r_{smax})^2 \quad (14)$$

The number of nodes required to cover $A$ for regular placement is equal to the number of hexagonal cells. For a typical scenario we consider the following system parameters: $R = 1000$ m, $r_{smax} = 50$ m, and $f_a = 90.69\%$. The approximate number of hexagonal cells required to cover the area of interest is 363.

The network coverage for random deployment has already been studied in section III. The comparison between random and regular deployment has been presented in the next section to show the impact of node deployment technique on network coverage.

## V. RESULTS AND DISCUSSIONS

In this section we present the numerical results to show the impact of sensing models on network coverage. Fig. 4 shows the comparative study of network coverage versus number of nodes for different sensing models. For Elfes sensing model we assume $\gamma = 1$. Also, we assume that maximum sensing range of a node is limited to 50 m. We have shown coverage fraction for two different values of $\lambda$, $\sigma$ and $R_1$. Curve-a is obtained for Boolean sensing model. Curves-c and f are obtained for Elfes sensing model for $\lambda = 0.01$/m and 0.03/m respectively while $R_1 = 0$. It is clear from Fig. 4 that for higher value of $\lambda$, more number of nodes are required to provide a certain coverage fraction. Curves-b and d are obtained for shadow-fading sensing model for $\sigma = 2$ dB and 8 dB respectively. It is also clear from the study that fading parameter degrades the network coverage. Curve-e is obtained for Elfes sensing model for $R_1 = 10$ m, $\lambda = 0.03$/m. It is clear from the study that best coverage fraction is achieved for Boolean sensing model. The degradation in coverage fraction for Elfes sensing model arises due to uncertainty of detection in the sensing model. Also, it is clear that the deterministic sensing model provides better network coverage than the probabilistic sensing model.

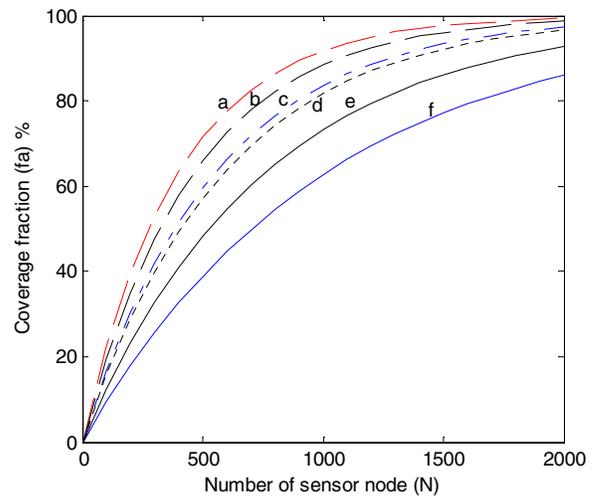

Fig. 4. Variation of coverage fraction for a circular area with radius 1000 m and maximum sensing radius $R_{max} = 50$ m: (a) Boolean sensing model; (b) Shadow fading sensing model ($\sigma = 2$ dB) [5]; (c) Elfes sensing model ($R_1 = 0$, $\lambda = 0.01$/m, $\gamma = 1$); (d) Shadow fading sensing model ($\sigma = 8$ dB) [5]; (e) Elfes sensing model ($R_1 = 10$ m, $\lambda = 0.03$/m, $\gamma = 1$); (f) Elfes sensing model ($R_1 = 0$, $\lambda = 0.03$/m, $\gamma = 1$)

Fig. 5 shows the variation of coverage fraction, $f_a$ with the normalized sensing radius, $r/r_{smax}$. It is clear that $f_a$ increases with the increase of $r$. However, the variation of $f_a$ is less for smaller value of $r/r_{smax}$. The maximum coverage fraction (90.69%) is achieved for regular placement when the normalized sensing ratio is equal to 1. It is clear from this study that 949 sensor nodes are required for random placement while regular placement demands only 363 nodes to achieve the coverage fraction 90.69%. This study implies regular placement demands less number of nodes to provide a given coverage fraction. However, the problem with regular placement is to place all the nodes at the specified positions which seem to be an infeasible situation in most of the cases. A sensor network may be deployed over an inaccessible terrain where the option is only random placement. Thus there is a trade-off between number of nodes and node placement strategy.

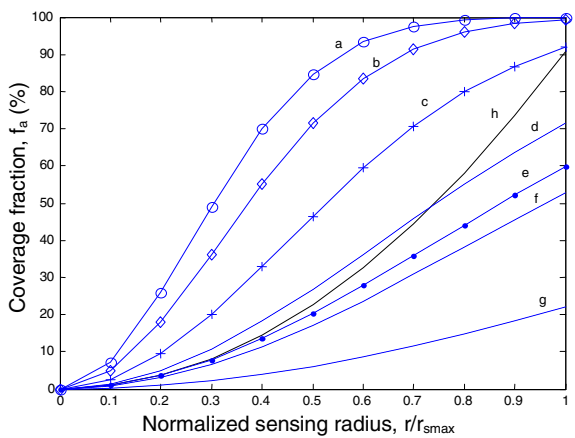

Fig. 5. Variation of coverage fraction, $f_a$ with normalized sensing radius for $R$ = 1000 m, $r_{smax}$ = 50 m: (a) – (g) Random placement and $N$ = 3000, 2000, 1000, 500, 363, 300 and 100; (h) Regular node placement according to Fig. 3

## VI. CONCLUSION

In this paper, the trade-off between number of nodes and node deployment strategy has been addressed. The regular placement results less number of nodes than the random placement to cover an area of interest. Analytical results are also provided to show the impact of sensing models on the network coverage. We have also investigated network coverage based on Poisson node distribution.

ACKNOWLEDGMENT

The authors would like to thank the anonymous reviewers for their valuable comments which helped to improve the quality of the paper.